\documentclass{aastex63}

\usepackage[T1]{fontenc}

\newcommand\Msun{M_\odot}
\newcommand\resc{r_{\rm esc}}

\received{June 1, 2019}
\revised{January 10, 2019}
\accepted{\today}
\submitjournal{ApJL}

\shorttitle{Fallback model}
\shortauthors{Ishizaki et al.}
\graphicspath{{./}{figures/}}
\begin{document}

\title{Fallback Accretion Model for the Years-to-Decades X-ray Counterpart to GW170817}

\correspondingauthor{Wataru Ishizaki}
\email{wataru.ishizaki@yukawa.kyoto-u.ac.jp}

\author[0000-0002-7005-7139]{Wataru Ishizaki}

\author[0000-0002-3517-1956]{Kunihito Ioka}

% \collaboration{1}{(AAS Journals Data Scientists collaboration)}

\affiliation{Center for Gravitational Physics, Yukawa Institute for Theoretical Physics, Kyoto University, Kyoto, 606-8502, Japan}

\author[0000-0003-4988-1438]{Kenta Kiuchi}
\affiliation{Center for Gravitational Physics, Yukawa Institute for Theoretical Physics, Kyoto University, Kyoto, 606-8502, Japan}
\affiliation{Max Planck Institute for Gravitational Physics (Albert Einstein Institute), Am M\"{u}hlenberg, Potsdam-Golm, 14476, Germany}



\begin{abstract}
A new component was reported in the X-ray counterpart to the binary neutron-star merger and gravitational wave event GW170817, exceeding the afterglow emission from an off-axis structured jet. The afterglow emission from the kilonova/macronova ejecta may explain the X-ray excess but exceeds the radio observations if the spectrum is the same. We propose a fallback accretion model that a part of ejecta from the neutron star merger falls back and forms a disk around the central compact object. In the super-Eddington accretion phase, the X-ray luminosity stays near the Eddington limit of a few solar masses and the radio is weak, as observed. This will be followed by a power law decay. The duration of the constant luminosity phase conveys the initial fallback timescale $t_0$ in the past. The current multi-year duration requires $t_0 > 3$--$30$ sec, suggesting that the disk wind rather than the dynamical ejecta falls back after the jet launch. Future observations in the next decades will probe the timescale of $t_0 \sim 10$--$10^4$ sec, around the time of extended emission in short gamma-ray bursts. The fallback accretion has not been halted by the $r$-process heating, implying that fission is weak on the year scale.
We predict that the X-ray counterpart will disappear in a few decades due to the $r$-process halting or the depletion of fallback matter.
\end{abstract}



\section{Introduction}\label{sec:intro}

Gravitational waves from the first binary neutron star merger event (BNS merger) GW170817 were observed by the Laser Interferometer Gravitational-Wave Observatory (LIGO) and the Virgo Consortium (LVC) \citep{2017PhRvL.119p1101A}.
This event was accompanied by a short-duration gamma-ray burst (sGRB) GRB 1780817A \citep{2017ApJ...848L..13A}.
The superluminal motion of the radio counterpart was detected, indicating that the relativistic jet was launched from the merged compact stars \citep{2018Natur.561..355M,2019Sci...363..968G} in the off-axis direction \citep[e.g.,][]{2018PTEP.2018d3E02I,2019MNRAS.487.4884I}.
In addition, a kilonova/macronova is also associated with GW170817 \citep{2017Natur.551...64A,2017ApJ...848L..19C,2017Sci...358.1556C,2017ApJ...848L..17C,2017Sci...358.1570D,2017Natur.551...80K,2017Sci...358.1559K,2017Sci...358.1583K,2017ApJ...848L..32M,2017ApJ...848L..18N,2017Sci...358.1574S,2017Natur.551...75S,2017ApJ...848L..16S,2017PASJ...69..102T,2017ApJ...848L..27T}.
From the modelling of the kilonova/macronova \citep[e.g.,][]{2017ApJ...848L..17C,2017Natur.551...80K,2017Sci...358.1559K,2017ApJ...851L..21V,2018ApJ...865L..21K}, it is suggested that the merger ejecta consists of at least two components: the dynamical ejecta and the post-merger ejecta.
The former is released promptly after coalescence in the dynamical timescale \citep[e.g.,][]{2013ApJ...773...78B,2013PhRvD..87b4001H,2016MNRAS.460.3255R,2015PhRvD..91f4059S,2016PhRvD..93l4046S,2017PhRvD..96l4005B,2017PhRvD..95b4029D}, and the latter is released in the secular viscous timescale \citep[e.g.,][]{2013MNRAS.435..502F,2015MNRAS.448..541J,2019MNRAS.482.3373F,2020ApJ...901..122F}. 

In GRB 170817A, an off-axis afterglow where an off-axis relativistic jet interacts with circumstellar medium (CSM), has been observed \citep{2017ApJ...848L..25H,2018ApJ...856L..18M,2017Natur.551...71T}.
The temporal evolution of the afterglow shows a much slower brightening than seen in top-hat jet afterglows, which is understood to be an afterglow from a structured jet \citep[e.g.,][]{2019ApJ...870L..15L,2019Sci...363..968G,2019MNRAS.489.1919T,2020MNRAS.493.3521B,2020MNRAS.497.1217T,2021MNRAS.501.5746T}.
The modeling of the afterglow has been successful, constraining not only the physical quantities of the jet (e.g., opening angle of the jet, jet luminosity, and structure), but also the density of CSM \citep[e.g.,][]{2018ApJ...856L..18M,2018ApJ...863L..18A,2019ApJ...870L..15L,2019MNRAS.489.1919T,2021MNRAS.501.5746T}.

Recently, the X-ray flux was observed several years after the coalescence \citep{2019ApJ...886L..17H,2020RNAAS...4...68H,2020GCN.29055....1H,2021GCN.29375....1H,2020MNRAS.498.5643T,2021arXiv210304821B}.
It has been reported that a 2--3$\sigma$ excess component over the prediction of the jet afterglow model \citep{2019ApJ...886L..17H,2021arXiv210304821B,2021GCN.29375....1H}.
\citet{2019ApJ...886L..17H} argued that this is a kilonova/macronova afterglow, which is generated by the interaction of the merger ejecta with CSM
\citep{2011Natur.478...82N,2014MNRAS.437L...6K,2014PhRvD..89f3006T,2018ApJ...852..105A,2018ApJ...867...95H,2019MNRAS.487.3914K,2021MNRAS.502.1843N}.
This model expects an increase of synchrotron emission also in the radio band.
However, no such radio re-brightening has been observed \citep{2019ApJ...886L..17H,2021arXiv210304821B}.
This may imply a different spectral index.

\begin{figure*}[t!]
	\centering
	\includegraphics[width=1.0\linewidth]{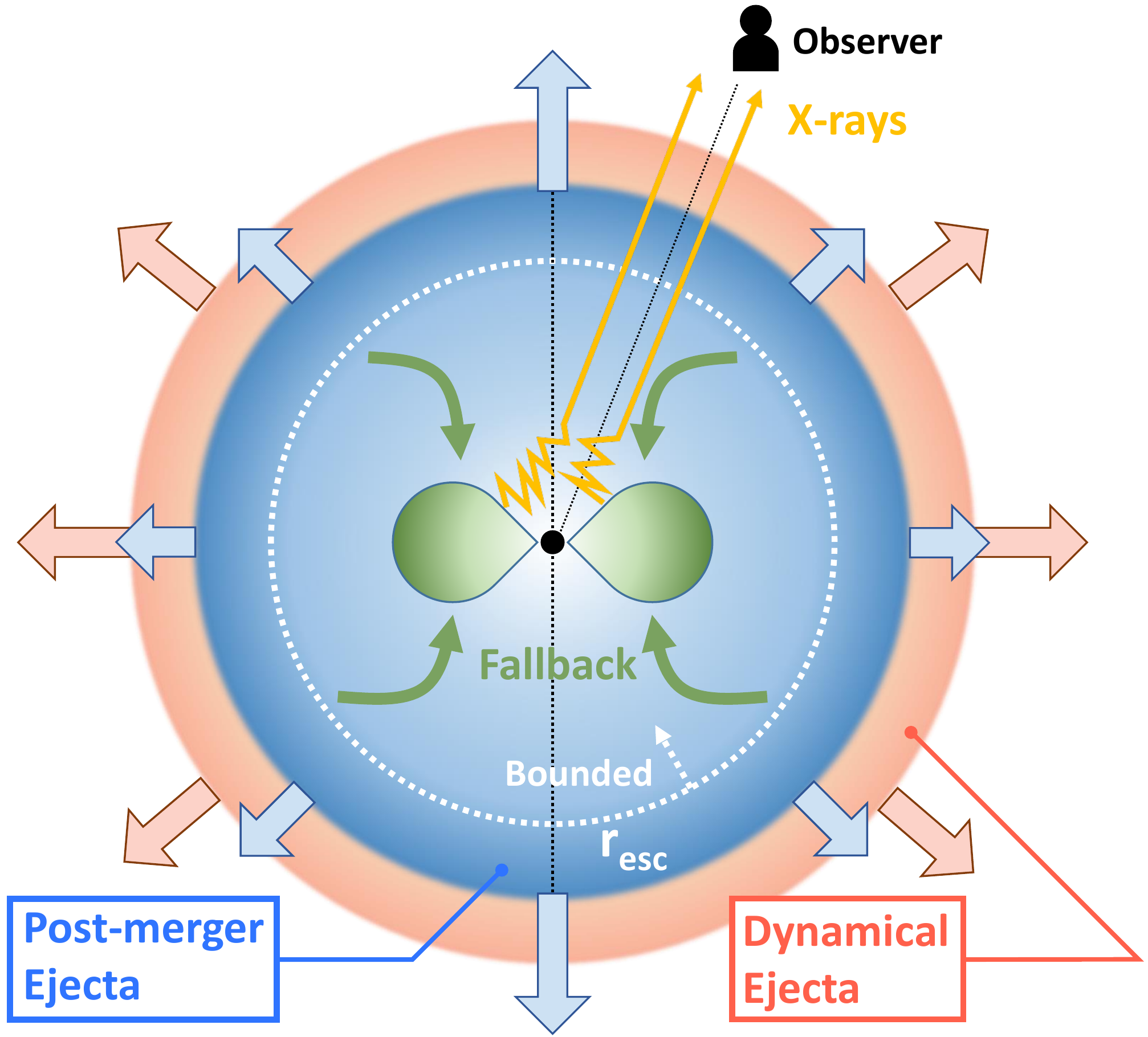}
	\caption{
	A schematic picture of our model.
    The post-merger ejecta (shown in blue) is blown out after the dynamical ejecta (shown in red) which released immediately after the compact binary coalescence.
    The post-merger ejecta inside a radius $r_{\rm esc}$ is still gravitationally bounded and will fallback into the central compact object.
    Since the fallback ejecta has finite angular-momentum, an accretion disk (shown in green) is formed around the central object.
    Photons from the accretion disk are mainly observed as X-rays (shown in orange).
	}
	\label{fig:manga}
\end{figure*}

In this Letter, we present an alternative interpretation for the X-ray excess in GW170817 (see Figure \ref{fig:manga}).
As shown by many theoretical studies, a part of the ejecta from the BNS merger is inevitably gravitationally bounded, regardless of whether they are dynamical or post-merger ejecta.
Since the mass $M$ of the central object of GW170817 is estimated as $M=2.74_{-0.01}^{+0.04}~\Msun$ \citep{2017PhRvL.119p1101A}, its Eddington luminosity is as follows:
%%%%%%%%%%%%%%%%%%%%%%%%%%%%%%%%%%%%%%%%%%%%%%%
\begin{equation}\label{eq:Ledd}
L_{\mathrm{edd}}=\frac{4 \pi G M m_{\mathrm{p}} c}{\sigma_{\mathrm{T}}} \sim 3.4 \times 10^{38} ~\mathrm{erg} ~\mathrm{s}^{-1}\left(\frac{M}{2.7 M_{\odot}}\right),
\end{equation}
%%%%%%%%%%%%%%%%%%%%%%%%%%%%%%%%%%%%%%%%%%%%%%%
where $m_p$ is the mass of the proton and $\sigma_{\rm T}$ is the Thomson scattering cross section.
This is consistent with the X-ray excess luminosity within a factor, implying that the X-ray excess could come from an accretion disk near the central compact object.
However, it is not clear whether the fallback accretion in the BNS merger can maintain a super-Eddington accretion for a few years or not\footnote{In fact, \citet{2018ApJ...856L..18M} applied the fallback accretion model to the early X-ray emission from the jet afterglow and concluded that the emission from the accretion flow never dominates the X-ray emission from GW170817. However, the parameter they adopted corresponds to the fallback of the dynamical ejecta (the mass acretion rate, $\dot{M}(t=1~\sec) \sim 10^{-3} \Msun~{\rm s}^{-1}$), which is not appropriate for considering light curves on an annual scale, as we will see later.}.
Thus, in this Letter, by constructing a X-ray light curve model from fallback accretion, we investigate whether the re-brightening is possible or not.
We also constrain the $r$-process heating rate from the X-ray light curve by applying the semi-analytic model of the mass accretion inhibited by radioactive heating due to the $r$-process elements developed in \citet{Ishizaki21a}.

\section{Fallback rate of the ejecta} \label{sec:EjModel}

%%%%%%%%%%%%%%%%%%%%%%%%
%%%%%%%%%%%%%%%%%%%%%%%

In this section, we estimate the fallback accretion rate for each of the two types of ejecta: the dynamical ejecta and the post-merger ejecta.
The accretion rate of dynamical ejecta, which is released promptly in the dynamical timescale after the merger when neutron stars are destroyed by tidal forces or the impact of the collision, can be modeled based on the numerical simulations.
On the other hand, the fallback rate of the post-merger ejecta, which is the subsequent disk wind launched mainly by viscous heating in the accretion disk in the secular timescale, is theoretically unclear, because the numerical relativity simulations of BNS mergers have not been calculated long enough to capture this fallback accretion \citep[see][for the best effort]{2020ApJ...901..122F}.
Therefore, for the post-merger ejecta, we estimate the mass accretion rate by constructing a simple model.

The fallback accretion rate decays in proportion to $t^{-5/3}$ after some initial time \citep{1988Natur.333..523R,1988Natur.333..644M}.
Thus we write the mass accretion rate as the following simple power-law form,
%%%%%%%%%%%%%%%%%%%%%%%%%%%%%%%%%%%%%%%%%%%%%%%
\begin{equation}\label{eq:mdot_model}
\dot{M}(t)=\frac{2}{3} \frac{M_{\rm tot,fb}}{t_0}\left(\frac{t}{t_0}\right)^{-5/3},
\end{equation}
%%%%%%%%%%%%%%%%%%%%%%%%%%%%%%%%%%%%%%%%%%%%%%%
where $t_0$ is the initial fallback time when the fallback accretion starts,
and $M_{\rm tot,fb}$ is the total fallback mass of the ejecta, which satisfies 
%%%%%%%%%%%%%%%%%%%%%%%%%%%%%%%%%%%%%%%%%%%%%%%
\begin{equation}
M_{\rm tot,fb}=\int_{t_{0}}^{\infty} \dot{M}(t) dt.
\end{equation}
%%%%%%%%%%%%%%%%%%%%%%%%%%%%%%%%%%%%%%%%%%%%%%%
In \citet{Ishizaki21a}, the accretion rate is calculated based on the profile of the dynamical ejecta derived from the numerical relativity simulation in \citet{2017PhRvD..96h4060K}.
The fallback accretion rate of the dynamical ejecta with zero temperature calculated using the same method as in \citet{Ishizaki21a}
is fitted by Equation (\ref{eq:mdot_model}) with $M_{\rm tot,fb}=4.0\times 10^{-3} \Msun$ and $t_0=4~{\rm msec}$.
We find that the simple power-law model describes well the expected accretion rate after $t>1~\sec$.

% On the other hand, the fallback rate of the post-merger ejecta, which is the subsequent disk wind mainly caused by viscous heating in the accretion disk, is theoretically unclear, because the numerical relativity simulations of BNS mergers have not been calculated long enough to capture these fallback accretion.
As mentioned earlier, the fallback accretion rate for the post-merger ejecta is theoretically unclear.
%However, the overall property of post-merger ejecta are relatively well understood, and it has been shown by 
The state-of-the-art study by \citet{2020ApJ...901..122F} showed that the post-merger ejecta mass is about $0.05$--$0.2~\Msun$ and its velocity is about $v=0.01c$--$0.1c$. 
\citet{2020arXiv201214711K} calculated the long term evolution of the post-merger ejecta using the result of \citet{2020ApJ...901..122F} as initial conditions.
While the disk outflow is blowing ($t\lesssim 8$ sec), the resultant angle-averaged mass density profile of the ejecta is approximately proportional to the power-law of a radial distance $r^{-2.4}$ for the ejecta with velocity less than about $0.04c$ (\citet{2020arXiv201214711K}, private communication, see also \citet{2020ApJ...901..122F}).
Hence, we assume that the mass density distribution of the post-merger ejecta is proportional to $r^{-2.4}$.

In order to estimate the fallback mass for the post-merger ejecta, we calculate the boundary radius between the gravitationally bounded and unbounded parts.
At this radius $\resc$, the kinetic and gravitational energies are balanced.
If the ejecta at $\resc$ moves under the energy conservation law, the radius $\resc$ evolves in time according to the following differential equation:
%%%%%%%%%%%%%%%%%%%%%%%%%%%%%%%%%%%%%%%%%%%%%%%
\begin{equation}
\frac{1}{2}\left(\frac{dr_{\rm esc}}{dt}\right)^2-\frac{GM}{r_{\rm esc}}=0.
\end{equation}
%%%%%%%%%%%%%%%%%%%%%%%%%%%%%%%%%%%%%%%%%%%%%%%
Integrating this over time and evaluating it at $t=t_0$, we obtain
%%%%%%%%%%%%%%%%%%%%%%%%%%%%%%%%%%%%%%%%%%%%%%%
\begin{equation}
%% r_{\mathrm{esc}}=\left(2 G M t_{0}^{2}\right)^{1/3},
r_{\mathrm{esc}}=\left(\frac{9}{2} G M t_{0}^{2}\right)^{1/3},
\end{equation}
%%%%%%%%%%%%%%%%%%%%%%%%%%%%%%%%%%%%%%%%%%%%%%%
where we assume that $\resc$ at $t_0$ is sufficiently large relative to the initial position.
At $t=t_0$, the mass inside $r=\resc$ is the total mass to fall back $M_{\rm tot,fb}$.
Given the mass density distribution $r^{-2.4}$,
the ratio of the fallback mass to the total mass $M_{\rm ej}$ of the post-merger ejecta, $f_{\rm fb}$, can be calculated as
%%%%%%%%%%%%%%%%%%%%%%%%%%%%%%%%%%%%%%%%%%%%%%%
\begin{equation}\label{eq:ffb}
f_{\rm fb}=\frac{M_{\rm tot,fb}}{M_{\rm ej}}=\frac{\int_{0}^{r_{\rm esc}}\rho_{\rm esc}\left(r/r_{\rm esc}\right)^{-2.4}4\pi r^2dr}{\int_{0}^{v_{\rm ej}t_{\rm 0}}\rho_{\rm esc}\left(r/r_{\rm esc}\right)^{-2.4}4\pi r^2dr}
=\left(\frac{r_{\rm esc}}{v_{\rm ej}t_{\rm 0}}\right)^{0.6},  
\end{equation}
%%%%%%%%%%%%%%%%%%%%%%%%%%%%%%%%%%%%%%%%%%%%%%%
where $\rho_{\rm esc}$ is the mass density at $r=\resc$ and $v_{\rm ej}$ is the velocity of the post-merger ejecta.
Note that the mass $(1-f_{\rm fb})M_{\rm ej}$ that does not fall back radiates the kilonova/macronova.
The numerical evaluation of the quantities in the post-merger ejecta is as follows:
%%%%%%%%%%%%%%%%%%%%%%%%%%%%%%%%%%%%%%%%%%%%%%%
\begin{equation}
r_{\rm esc}\sim3.43\times10^{9}~{\rm cm}
 ~\left(\frac{t_{0}}{5~{\rm sec}}\right)^{2/3}
 ~\left(\frac{M}{2.7~M_{\odot}}\right)^{1/3},
\end{equation}
%%%%%%%%%%%%%%%%%%%%%%%%%%%%%%%%%%%%%%%%%%%%%%%
%%%%%%%%%%%%%%%%%%%%%%%%%%%%%%%%%%%%%%%%%%%%%%%
\begin{equation}\label{eq:ffb_num}
f_{\rm fb}\sim0.72
 ~\left(\frac{t_{0}}{5~{\rm sec}}\right)^{-0.2}
 ~\left(\frac{M}{2.7~M_{\odot}}\right)^{0.2}
 ~\left(\frac{v_{\rm ej}}{0.04c}\right)^{-0.6},
\end{equation}
%%%%%%%%%%%%%%%%%%%%%%%%%%%%%%%%%%%%%%%%%%%%%%%
%%%%%%%%%%%%%%%%%%%%%%%%%%%%%%%%%%%%%%%%%%%%%%%
\begin{equation}\label{eq:pmej_Mtot}
M_{\rm tot,fb}
%=f_{\rm fb}M_{\rm ej}
\sim
5.72\times 10^{-2}~M_{\odot}
 ~\left(\frac{t_{0}}{5~{\rm sec}}\right)^{-0.2}
 ~\left(\frac{M}{2.7~M_{\odot}}\right)^{0.2}
 ~\left(\frac{M_{\rm ej}}{0.08~M_\odot}\right)
 ~\left(\frac{v_{\rm ej}}{0.04c}\right)^{-0.6}.
\end{equation}
%%%%%%%%%%%%%%%%%%%%%%%%%%%%%%%%%%%%%%%%%%%%%%%
The time $t_0$ at which the fallback starts in the post-merger ejecta is for example the time when the outflow from the central object stops blowing and the fallback cannot be supported by the ram pressure of the outflow.
However, because the theoretical understanding of the fallback accretion of the post-merger ejecta is not enough as mentioned above, the exact value of $t_0$ is not known.
Therefore we treat $t_0$ as a free parameter in this study.
For the mass $M_{\rm ej}$ and velocity $v_{\rm ej}$ of the ejecta, we adopt the values that are consistent with the kilonova/macronova light curve used in \citet{2020arXiv201214711K}, who properly treat the radiation transfer across the different ejecta components.

\section{X-ray light curve of the fallback accretion in BNS merger} \label{sec:XLC}

The ejecta that falls back onto the central object forms an accretion disk, and emits mainly X-rays as will be discussed later.
Here, we assume that the central object collapses to a black hole (BH) with mass $M=2.7 \Msun$ at least at the time $t\sim 1$ yr.
These X-rays are only observable if they are not absorbed by the kilonova/macronova ejecta in the line of sight \citep{2016ApJ...818..104K,2018ApJ...854...60M,2018ApJ...856L..18M}.
According to \citet{2018ApJ...854...60M}, the time required for the dynamical ejecta to become optically thin is estimated as follows:
%%%%%%%%%%%%%%%%%%%%%%%%%%%%%%%%%%%%%%%%%%%%%%%%
\begin{equation}\label{eq:tthin}
t_{\rm thin}
=\left(\frac{3\hat{K}_{\rm X}M_{\rm ej}}{4\pi v_{\rm ej}^2}\right)^{1/2}
\sim
0.65 {\rm ~yr}
\left(\frac{\hat{K}_{\mathrm{X}}}{100 \mathrm{~cm}^{2} \mathrm{~g}^{-1}}\right)^{1 / 2}
\left(\frac{M_{\rm ej}}{0.08~M_{\odot}}\right)^{1 / 2}
\left(\frac{v_{\rm ej}}{0.1 c}\right)^{-1},
\end{equation}
%%%%%%%%%%%%%%%%%%%%%%%%%%%%%%%%%%%%%%%%%%%%%%%%
where $\hat{K}_{\mathrm{X}}$ is the typical opacity for a 10 keV X-ray photon.
For lower energy X-ray photons, $\hat{K}_{\mathrm{X}}$ is much larger, and the expected bound-free opacity of neutral or singly ionized heavy $r$-process nuclei is about $\hat{K}_{\mathrm{X}}\sim 10^3 \mathrm{~cm}^{2} \mathrm{~g}^{-1}$ \citep[e.g.,][]{2017LRR....20....3M,2018ApJ...856L..18M}.
In Equation (\ref{eq:tthin}), the value is evaluated for the post-merger ejecta, because the dynamical ejecta has less mass and higher velocity than the post-merger ejecta.
For the ejecta mass, we adopt $M_{\rm ej}=0.08~\Msun$, which is the mass of the wind component based on the \citet{2020arXiv201214711K}.
This value is consistent with the observation-based phenomenological model of kilonova/macronova \citep[e.g.,][]{2017ApJ...848L..19C,2017ApJ...848L..17C,2017Sci...358.1559K,2017ApJ...851L..21V}, since a large fraction of the mass $M_{\rm rj}$ fallbacks and does not contribute to the kilonova/macronova (see Equation (\ref{eq:ffb}) and (\ref{eq:ffb_num})).
For the ejecta velocity, we adopt $v_{\rm ej} = 0.1c$ not $0.04c$ to evaluate Equation (\ref{eq:tthin}).
The geometric configuration of GW170817 is such that the line of sight direction is close to the axial direction (i.e., the viewing angle $\theta_{\rm v}$ is about $\theta_{\rm v}\sim 25$ deg \citep{2018Natur.561..355M}).
According to \citet{2020arXiv201214711K}, ejecta on the polar direction are about twice as fast as those in the equatorial plane, so that we evaluate them with the parameter $v_{\rm ej} = 0.1c$.
On the other hand, if the viscous parameter $\alpha\sim \mathcal{O}(0.01)$, since the post-merger ejecta is expected to have a larger electron fraction \citep[see][]{2020ApJ...901..122F} than that of the dynamical ejecta, heavy $r$-process nuclei would be less formed and then the value of $\hat{K}_{\mathrm{X}}$ would be smaller (For details, see the mass energy-transfer coefficient data provided by NIST\footnote{https://www.nist.gov/pml/x-ray-mass-attenuation-coefficients}).
Then, after a year, the kilonova/macronova ejecta is expected to become optically thin to X-rays.

The fallback ejecta converts its gravitational energy into radiation by viscous heating in the accretion disk.
Initially the mass accretion rate is much larger than the Eddington accretion rate.
In this super-Eddington accretion phase, the luminosity is roughly limited by the Eddington luminosity $L_{\rm edd}$.
Many theoretical calculations of the radiative transport in super-Eddington accretion show that the isotropic bolometric luminosity seen by slightly off-axis observers (i.e., for GW170817, $\theta_{\rm v} \sim 25$ deg) is about twice that of $L_{\rm edd}$ \citep[e.g.,][]{2005ApJ...628..368O,2016MNRAS.456.3929S,2017PASJ...69...33O,2018MNRAS.479.3936A,2021arXiv210111028K}.
Hence, we estimate the bolometric luminosity in the super-Eddington phase as follows:
%%%%%%%%%%%%%%%%%%%%%%%%%%%%%%%%%%%%%%%%%%%%%%%
\begin{equation}\label{eq:Lxint_sup}
L_{\rm bol}=f_{\rm edd}L_{\rm edd},
\end{equation}
%%%%%%%%%%%%%%%%%%%%%%%%%%%%%%%%%%%%%%%%%%%%%%%
where $f_{\rm edd}$ is a constant and we adopt $f_{\rm edd}=2$.
As the mass accretion rate becomes small, the radiative transfer effect becomes less effective and the radiative efficiency increases.
For the mass accretion rate smaller than the Eddington accretion rate, the bolometric luminosity can be estimated as follows:
%%%%%%%%%%%%%%%%%%%%%%%%%%%%%%%%%%%%%%%%%%%%%%%
\begin{equation}\label{eq:Lxint_sub}
L_{\rm bol}=\eta \dot{M} c^2
\end{equation}
%%%%%%%%%%%%%%%%%%%%%%%%%%%%%%%%%%%%%%%%%%%%%%%
where $\eta$ is the radiative efficiency and we adopt $\eta=0.1$ \citep[e.g.,][]{2005ApJ...628..368O}.
% We consider the following simple model using Equations (\ref{eq:Lxint_sup}) and (\ref{eq:Lxint_sub}) as an X-ray light curve model which roughly incorporates the change in radiative efficiency with the time evolution of the mass accretion rate.
Combining Equations (\ref{eq:Lxint_sup}) and (\ref{eq:Lxint_sub}), we estimate the luminosity with the following simple formula:
%%%%%%%%%%%%%%%%%%%%%%%%%%%%%%%%%%%%%%%%%%%%%%%
\begin{equation}\label{eq:Lx}
L_{\rm bol}=\min\left(\eta \dot{M} c^2,f_{\rm edd}L_{\rm edd}\right).
\end{equation}
%%%%%%%%%%%%%%%%%%%%%%%%%%%%%%%%%%%%%%%%%%%%%%%
Note that the isotropic luminosity already accounts for the beaming effect.

% In order to convert the bolometric luminosity $L_{\rm bol}$ to the observed X-ray luminosity $L_{\rm X}$ in Chandra band ($0.3$--$10$ keV), we estimate the conversion factor $f_{\rm X}$ from theoretical calculations of the emission spectrum of super-Eddington accretion flow.

In order to convert the bolometric luminosity $L_{\rm bol}$ to the observed X-ray luminosity $L_{\rm X}$ in Chandra band ($0.3$--$10$ keV), first of all, we make a rough estimate of the spectrum of the disk emission.
By using equation (\ref{eq:Lxint_sup}), we can estimate the effective temperature at the innermost stable circular orbit $R=\xi R_{\rm s}=3 \xi_3 R_{\rm s}$ as
%%%%%%%%%%%%%%%%%%%%%%%%%%%%%%%%%%%%%%%%%%%%%%%
\begin{equation}
T_{\rm eff,max}=\left[\frac{f_{\rm edd}L_{\rm edd}}{4\pi \left(\xi R_{\rm s}\right)^2 \sigma}\right]^{1/4}
\sim
1.50 %1.5029
~{\rm keV}~\xi_3^{-1/2}~f_{{\rm edd},2}^{1/4}~\left(\frac{M}{2.7~M_\odot}\right)^{-1/4},
\end{equation}
%%%%%%%%%%%%%%%%%%%%%%%%%%%%%%%%%%%%%%%%%%%%%%%
where $f_{\rm edd,2}=f_{\rm edd}/2$, $R_{\rm s}$ is the Schwarzschild radius of the central object, and $\sigma$ is the Stefan–Boltzmann constant.
For the disk in the super-Eddington phase, the effective temperature depends on the radius as $T_{\rm eff}\propto R^{-1/2}$ \citep{2000PASJ...52..133W}.
If we assume that the emission at each radius is the blackbody radiation at the effective temperature $T_{\rm eff}$, the emission spectrum $L_\nu$ of the disk can be estimated as follows:
%%%%%%%%%%%%%%%%%%%%%%%%%%%%%%%%%%%%%%%%%%%%%%%
\begin{equation}
L_\nu=2\int_{\xi R_{\rm s}}^{R_{\rm max}}2\pi RB_\nu\left(T_{\rm eff}(R)\right) dR 
\propto R^2T_{\rm eff}^3
\propto \nu^{-1},
\end{equation}
%%%%%%%%%%%%%%%%%%%%%%%%%%%%%%%%%%%%%%%%%%%%%%%
where $R_{\rm max}$ is the outer radius of the disk, $B_\nu$ is the blackbody spectrum, and, here we used the fact that the peak of blackbody radiation is proportional to $T_{\rm eff}^3$ and Wien's displacement law $\nu\propto T_{\rm eff}$.
Hence, the conversion factor can be calculated as follows:
%%%%%%%%%%%%%%%%%%%%%%%%%%%%%%%%%%%%%%%%%%%%%%%
\begin{equation}
f_{\rm X}=\frac{\int_{0.3~{\rm keV}}^{10~{\rm keV}} L_\nu d\nu}{L_{\rm bol}}\sim \frac{\log\left[\min({T_{\rm eff,max},10~{\rm keV}})\right]-\log\left[\max({T_{\rm eff,min},0.3~{\rm keV}})\right]}{\log(T_{\rm eff,max})-\log(T_{\rm eff,min})},
\end{equation}
%%%%%%%%%%%%%%%%%%%%%%%%%%%%%%%%%%%%%%%%%%%%%%%
where $T_{\rm eff,min}=T_{\rm eff,max}(R_{\rm max}/(\xi R_{\rm s}))^{-1/2}$ is the effective temperature at the outer radius.
Evaluating the value of $f_{\rm X}$ for $R_{\rm max}\lesssim 5000~{\rm km}$ gives $f_{\rm X}\sim 1$.
More precisely, in the super-Eddington accretion phase, the system is inevitably optically thick against absorption and scattering, so that we estimate $f_{\rm X}$ based on numerical calculations that incorporate the effect of radiative transfer.
According to the Monte-Carlo simulation based on hydrodynamic simulations of the super-Eddington flow, the spectrum with $10 \Msun$ BH as a central source has a peak at a few keV, not depending largely on the accretion rate, and has approximately $f_{\rm X}\sim 0.5$--$0.8$ \citep[e.g.,][]{2012ApJ...752...18K,2013ApJ...769..156S,2017MNRAS.469.2997N,2017PASJ...69...92K,2020arXiv201205386K}.
The spectral energy distributions (SEDs) obtained by these simulations show a flat frequency dependence centered at a few keV, corresponding to $\Gamma \sim 2$.
\citet{2019ApJ...886L..17H} reported a photon index of $\Gamma={1.23}_{-1.03}^{+1.05}$ for the X-ray emission two years after the merger, which is consistent with the X-ray from super-Eddington flow within the uncertainty.
\citet{2017PASJ...69...92K} investigated the dependence of the spectrum on the BH mass with $10$--$10^4~\Msun$, and found that the larger the BH mass, the broader the spectrum centered at a few keV.
This implies that the smaller the BH mass, the larger $f_{\rm X}$ tends to be.
Therefore, extrapolating the result of \citet{2017PASJ...69...92K} to the case for the BH mass $2.7 \Msun$, we adopt $f_{\rm X}=0.7$.
Furthermore, we should also take into account the X-ray absorption by ejecta.
Since the mean density of ejecta evolves in time as $\bar{\rho}\propto t^{-3}$, the temporal evolution of the optical depth $\tau$ can be evaluated as $\tau=\hat{K}_{\rm X} \bar{\rho} ct\propto t^{-2}$.
At $t=t_{\rm thin}$, the ejecta becomes optically thin (i.e. $\tau\sim1$), so that the observed X-ray luminosity is written as 
% Evaluating the amount of absorption using the timescale $t_{\rm thin}$ in Equation (\ref{eq:tthin}), the observed X-ray luminosity is written as 
%%%%%%%%%%%%%%%%%%%%%%%%%%%%%%%%%%%%%%%%%%%%%%%
\begin{equation}
% L_{\rm X}=f_{\rm X}\left[1-\exp\left(-t/t_{\rm thin}\right)\right]\min\left(\eta \dot{M} c^2,f_{\rm edd}L_{\rm edd}\right)
% L_{\rm X}=f_{\rm X}\left(1-e^{-t/t_{\rm thin}}\right)L_{\rm bol}.
L_{\rm X}=f_{\rm X}e^{-\left(t/t_{\rm thin}\right)^{-2}}L_{\rm bol}.
%\min\left(\eta \dot{M} c^2,f_{\rm edd}L_{\rm edd}\right)
\end{equation}
\begin{figure*}[t!]
	\centering
	\includegraphics[width=1.0\linewidth]{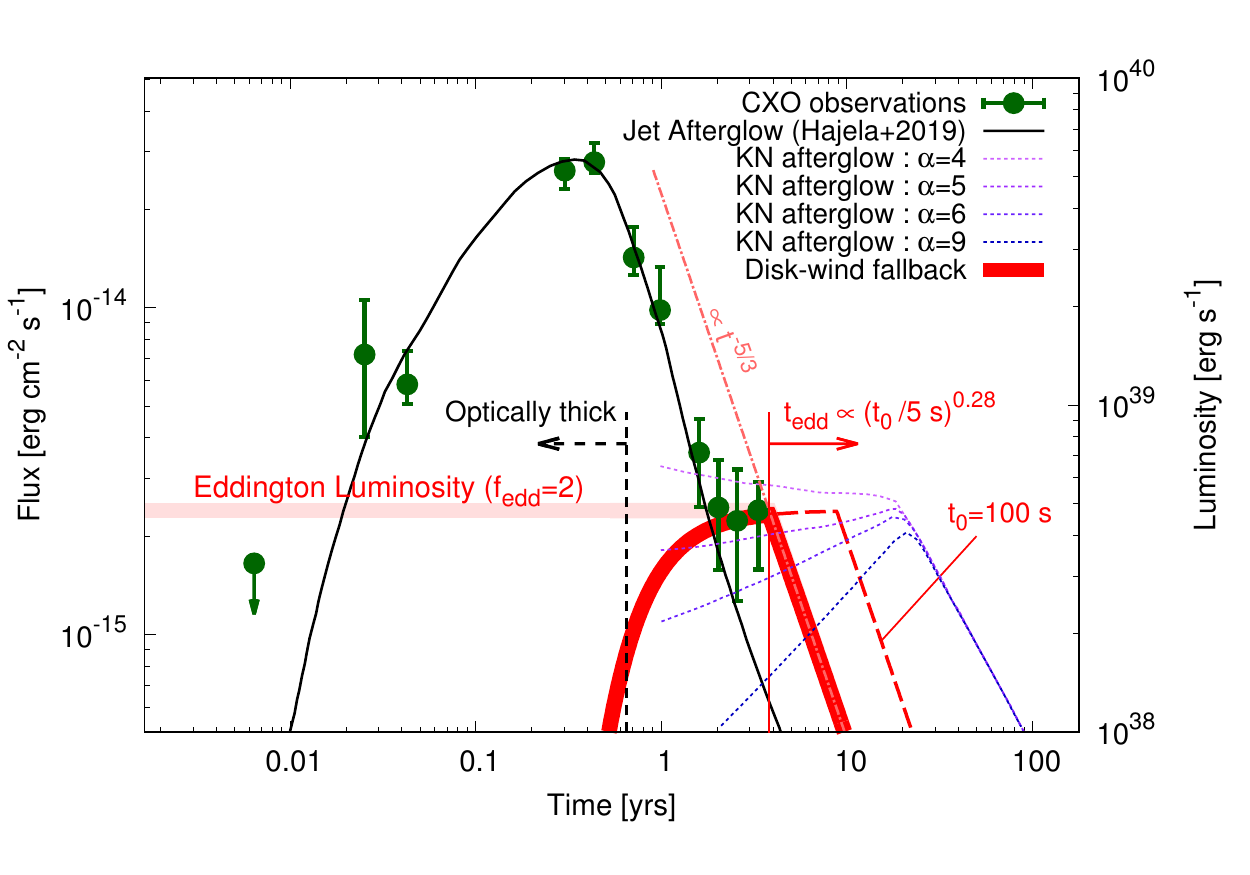}
	\caption{
	The red thick curve shows the X-ray flux from the fallback accretion of the post-merger ejecta (disk wind), and the light red thick curve represents the X-ray flux before the absorption (approximately the Eddington flux).
	The thin red dashed-and-dotted line is the power-law of time ($t^{-5/3}$), corresponding to the X-ray emission simply proportional to the mass accretion rate (i.e., ignoring the radiative transfer and the absorption).
    The black dashed thick line shows the time before which the ejecta is optically thick to the X-rays (see Equation (\ref{eq:tthin})).
    The parameters used in the calculation of the post-merger ejecta are the total mass of the ejecta, $M_{\rm ej} = 0.08 \Msun$, the velocity of the ejecta, $v_{\rm ej}=0.04c$, the mass of the central object, $M=2.7 \Msun$, and the initial fallback time, $t_0=5$ sec.
    The red long dashed line represents the case for $t_0=100~\sec$ with the other parameters being the same.
    The break of the thick curve corresponds to the transition time from the super-Eddington regime to the sub-Eddington regime (see Equation (\ref{eq:tedd_pm})).
    The green points are the results of the Chandra observations performed by \citet{2019ApJ...886L..17H,2020RNAAS...4...68H,2020GCN.29055....1H,2020MNRAS.498.5643T}.
    The thin black curve is the model curve for the jet afterglow by \citet{2019ApJ...886L..17H}.
    The thin dotted curves are the model curves of the X-ray emission from the kilonova/macronova afterglow by \citet{2019ApJ...886L..17H}, and the difference in color indicates the difference in density distribution of the fast tail.
	}
	\label{fig:lx}
\end{figure*}
%%%%%%%%%%%%%%%%%%%%%%%

\begin{figure*}[t!]
	\centering
	\includegraphics[width=1.0\linewidth]{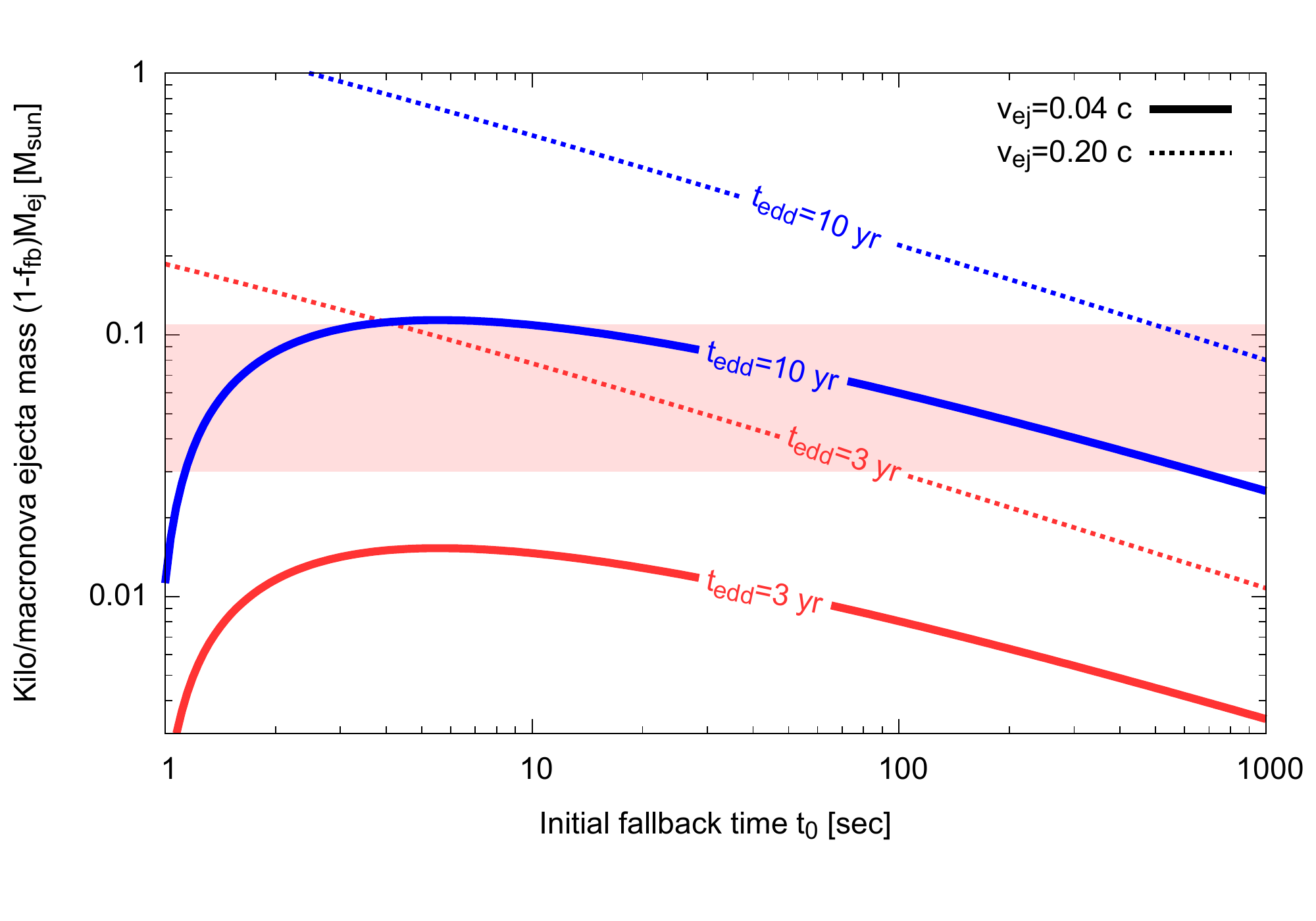}
	\caption{
	Contours of the timescale of the super-Eddington phase $t_{\rm edd}$ in the plane of the initial fallback time $t_0$ and the mass of the kilonova/macronova ejecta $(1-f_{\rm ej})M_{\rm ej}$.
	The red and blue curves show the cases of $t_{\rm edd}= 3 ~{\rm yr}$ and $t_{\rm edd}=10~{\rm yr}$, respectively.
    The thick and dashed curves show the cases of $v_{\rm ej}=0.04~c$ and $v_{\rm ej}=0.2 ~c$, respectively.
    The other parameters used in the calculation are the mass of the central object $M=2.7~\Msun$, the radiation efficiency $\eta=0.1$, and the correction factor in super-Eddington regime $f_{\rm edd}=2$.
    The reason why the curves with $v_{\rm ej}=0.04 ~c$ turn around $t_0\lesssim5~\sec$ is because $f_{\rm fb} \sim 1$.
    The red shaded region represents the range of the kilonova/macronova ejecta mass obtained by several observation-based phenomenological modeling \citep[e.g.,][]{2017ApJ...848L..19C,2017ApJ...848L..17C,2017Sci...358.1559K,2017ApJ...851L..21V,2020arXiv201214711K}.
    Note, however, that there is a range of $0.04c$--$0.3c$ in the ejecta velocity in these models.
    % \kkcom{Maybe better to plot the region of the kilnova ejecta suggested by GW170817 observation.}
	}
	\label{fig:mkn}
\end{figure*}

In Figure \ref{fig:lx}, the X-ray luminosity associated with the fallback accretion of the post-merger ejecta in Equation (\ref{eq:Lx}) is shown by the red curve.
The dashed line indicates the time before which the kilonova/macronova ejecta is optically thick and the X-rays are absorbed.
The X-ray light curve shows a constant luminosity up to a certain time $t_{\rm edd}$, after which it declines in proportion to $t^{-5/3}$.
The time $t_{\rm edd}$ corresponds to the time when Equations (\ref{eq:Lxint_sup}) and (\ref{eq:Lxint_sub}) are equal.
For the dynamical ejecta, we have
%%%%%%%%%%%%%%%%%%%%%%%%%%%%%%%%%%%%%%%%%%%%%%%
\begin{equation}\label{eq:tedd_dy}
t_{\rm edd}\sim 16~{\rm days}
~\eta_{0.1}^{3/5}
~f_{{\rm edd},2}^{-3/5}
~\left(\frac{t_{0}}{4~{\rm msec}}\right)^{2/5}
~\left(\frac{M}{2.7~M_{\odot}}\right)^{-3/5}
~\left(\frac{M_{\rm tot,fb}}{4\times 10^{-3}~M_\odot}\right)^{3/5}.
\end{equation}
%%%%%%%%%%%%%%%%%%%%%%%%%%%%%%%%%%%%%%%%%%%%%%%
For the post-merger ejecta, we calculate using Equations (\ref{eq:mdot_model}) and (\ref{eq:pmej_Mtot}) as follows:
%%%%%%%%%%%%%%%%%%%%%%%%%%%%%%%%%%%%%%%%%%%%%%%
\begin{equation}\label{eq:tedd_pm}
% t_{\rm edd}\sim 1395~{\rm days}
% ~\eta_{0.1}^{3/5}
% ~f_{{\rm edd},2}^{-3/5}
% ~\left(\frac{t_{0}}{5~{\rm sec}}\right)^{1/5}
% ~\left(\frac{M}{2.7~M_{\odot}}\right)^{-2/5}
% ~\left(\frac{M_{\rm ej}}{0.08~M_\odot}\right)^{3/5}
% ~\left(\frac{v_{\rm ej}}{0.04c}\right)^{-3/5}.
t_{\rm edd}\sim 1395~{\rm days}
~\eta_{0.1}^{3/5}
~f_{{\rm edd},2}^{-3/5}
~\left(\frac{t_{0}}{5~{\rm sec}}\right)^{0.28}
~\left(\frac{M}{2.7~M_{\odot}}\right)^{-0.48}
~\left(\frac{M_{\rm ej}}{0.08~M_\odot}\right)^{3/5}
~\left(\frac{v_{\rm ej}}{0.04c}\right)^{-0.36}.
\end{equation}
%%%%%%%%%%%%%%%%%%%%%%%%%%%%%%%%%%%%%%%%%%%%%%%
As can be seen from Equation (\ref{eq:tedd_dy}), the X-rays from the fallback of the dynamical ejecta is absorbed by the ejecta on a timescale earlier than a year, and this cannot be the main component explaining the X-ray excess.
On the other hand, the X-rays associated with the fallback of the post-merger ejecta can explain the X-ray excess.
Figure \ref{fig:mkn} shows the relationship between the initial fallback time $t_0$ and the mass of the kilonova/macronova ejecta $(1-f_{\rm fb})M_{\rm ej}$, for a given $t_{\rm edd}$.
The longer $t_{\rm edd}$, the more kilonova/macronova ejecta mass is required for the same $t_0$.
Note that for $v_{\rm ej}=0.04~c$, at $t_0 \lesssim 5 \sec$, $f_{\rm fb}$ is roughly about unity, so that the mass of the kilonova/macronova ejecta is small, but the total ejecta mass itself is very large.

Our model assumes that the mass accretion rate evolves in time as $t^{-5/3}$, but there can be variations in the actual temporal evolution \citep[e.g.,][]{1990ApJ...351...38C,1999MNRAS.306L...9O,2008MNRAS.388.1729K,2008MNRAS.390..781M}.
For example, for the radiatively inefficient disk, the mass accretion rate decreases in proportion to $t^{-4/3}$.
Generally, the temporal evolution of the light curve diminishing depends on the mass accretion rate and the mass of the accretion disk.
Here, for simplicity, we assume and continue to use equations (\ref{eq:mdot_model}) and (\ref{eq:Lx}) in this letter.
Furthermore, equation (\ref{eq:tedd_pm}) suggests that the system continues to be in a super-Eddington phase until now, which implies that the outflow is ejected from the radiatively inefficient disk.
The effect of the outflow on the radiative transfer has already been incorporated in the evaluation of $f_{\rm edd}$ and $f_{\rm X}$ based on the results of numerical simulations.
Some of the outflow can interact with the kilonova/macronova ejecta to form shock and release their kinetic energy.
However, the outflow that is fast enough to catch up with the kilonova/macronova ejecta is limited to those that escape from the fairly inner edge of the disk.

%%%%%%%%%%%%%%%%%%%%%%%%
%%%%%%%%%%%%%%%%%%%%%%%

\section{Implication to \textit{r}-process heating} \label{sec:rprocess}
\citet{Ishizaki21a} showed that the mass accretion in fallback is suppressed by the heating due to the decay of $r$-process elements.
According to \citet{Ishizaki21a}, the mass accretion rate decreases to 10\% compared to the usual power-law behavior $t^{-5/3}$ (see Equation (\ref{eq:mdot_model})) on a timescale called the halting time $t_{\rm halt}$, which is determined as the solution to the following equation,
%%%%%%%%%%%%%%%%%%%%%%%%%%%%%%%%%%%%%%%%%%%%%%%
\begin{equation}\label{eq:halting_time_def}
t_{\rm halt}=K\left[\frac{\left(GM\right)^2}{\dot{q}(t_{\rm halt})^3}\right]^{1/5},
\end{equation}
%%%%%%%%%%%%%%%%%%%%%%%%%%%%%%%%%%%%%%%%%%%%%%%
where $K\sim 2.6$ is a constant obtained from numerical calculations \citep[see][for details]{Ishizaki21a},
and $\dot{q}(t)$ is the radioactive heating rate per unit mass.
With the typical heating rate $\dot{q}\sim 10^{10}~{\rm erg}~{\rm g}^{-1}~{\rm s}^{-1}(t/1~{\rm day})^{-1.3}$, which is suggested by the nucleosynthesis calculations and the kilonova/macronova modelings \citep[e.g.,][]{2014ApJ...789L..39W,2017PASJ...69..102T}, the halting time can be estimated as
%%%%%%%%%%%%%%%%%%%%%%%%%%%%%%%%%%%%%%%%%%%%%%%
\begin{equation}\label{eq:halting_time_13}
% t_{\rm halt}\sim1.1\times 10^7~{\rm sec}
t_{\rm halt}\sim2.4\times 10^5~{\rm sec}
\left(\frac{M}{2.7~\Msun}\right)^{1.82}
\left(\frac{\dot{q}(t=1~{\rm day})}{10^{10}~{\rm erg}~{\rm g}^{-1}~{\rm s}^{-1}}\right)^{-2.72}.
\end{equation}
%%%%%%%%%%%%%%%%%%%%%%%%%%%%%%%%%%%%%%%%%%%%%%%
The halting time is sensitive to the uncertainty of the radioactive heating rate.
As reported by \citet{2020arXiv201011182B}, due to the uncertainty of nuclear physics, there is one order magnitude uncertainty of the heating rate at $\mathcal{O}(1)$ day.
The corresponding uncertainty of the halting time is a factor of $\sim 30$ up and down together (see Equation (\ref{eq:halting_time_13})).
Furthermore, after $\mathcal{O}(10)$ days of the coalescence, the gamma-rays emitted by radioactive decay can escape from the ejecta without ejecta heating.
For example, for the heating rate calculated by \citet{2014ApJ...789L..39W}, the effective heating rate drops to about half after $\mathcal{O}(10)$ days.
Note that the fraction of gamma-rays in radioactive heating is also affected by the uncertainty of nuclear physics.
Combining the effects of the insufficient thermalization and the uncertainty of the heating rate (assuming $0.5$ digit up and down even at $\mathcal{O}(10)$ years), the uncertainty of the halting time ranges about $10^4$--$10^8$ sec \citep[see][]{Ishizaki21a}.

%%%%%%%%%%%%%%%%%%%%%%%%
\begin{figure*}[t!]
	\centering
	\includegraphics[width=1.0\linewidth]{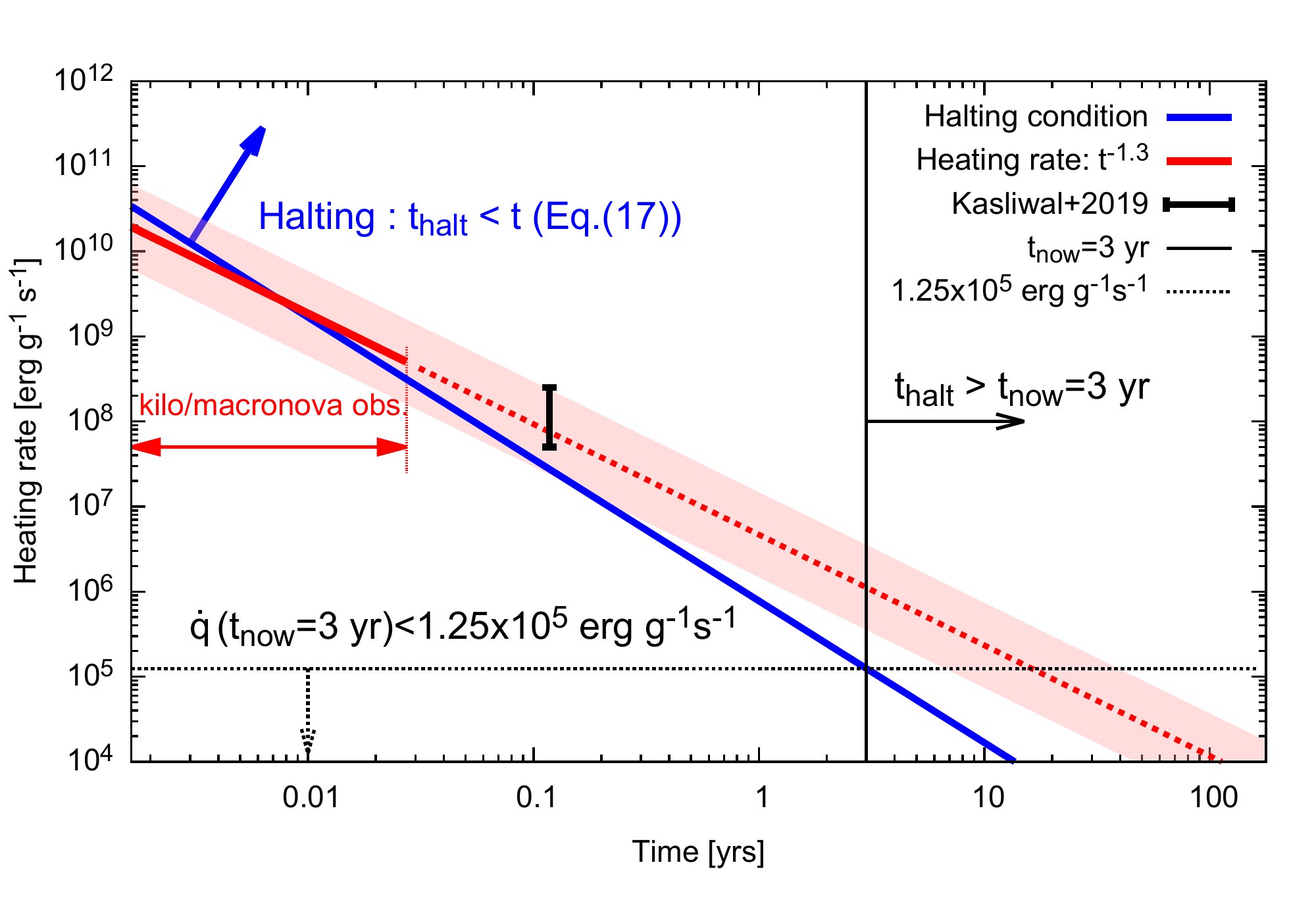}
	\caption{
	The blue line shows the condition above	which halting occurs, and if the heating rate is higher than this line, the mass accretion rate decreases rapidly (i.e., the halting).
    The red line is a simple power-law approximation of the radioactive heating rate, as suggested by the modeling of the kilonova/macronova of GW170817.
    The black dot with an error bar is the radioactive heating rate at $t=43$ day as obtained by \citet{2019MNRAS.tmpL..14K} using the mid-infrared observations by Spitzer.
    % If this simple power-law heating continues for $\mathcal{O}(10)$ days, the halting time will be the value shown in Equation (\ref{eq:halting_time_13}) (indicated by the red circle). 
    % \kkcom{There in no red circle in this revised figure.}
    In order for the halting not to have occurred, at the current time $t_{\rm now}=3$ yr (solid black line), the current heating rate must be less than $1.25\times 10^5 ~{\rm erg}~{\rm g}^{-1}~{\rm s}^{-1}$ (dashed black line).
	}
	\label{fig:halting}
\end{figure*}
%%%%%%%%%%%%%%%%%%%%%%%

After the halting time $t_{\rm halt}$, the mass accretion rate decreases more rapidly than $t^{-5/3}$.
If the heating rate can be written as a simple power-law form $\dot{q}\propto t^{-p}$, the temporal evolution of the mass accretion rate after the halting time can be written as
%%%%%%%%%%%%%%%%%%%%%%%%%%%%%%%%%%%%%%%%%%%%%%%
\begin{equation}\label{eq:mdot_analytic_decay}
\frac{\dot{M}}{\dot{M}_0}=\left(\frac{t}{t_{\rm halt}}\right)^{(3p-15)/10}
\exp\left[-f_{\rm PL}\frac{5}{5-3p}\frac{3\sqrt{2}}{\beta_0^3 K^{1/2}}\frac{f_1}{f_0}\left(1-\left(\frac{t_{\rm halt}}{t}\right)^{\frac{5-3p}{10}}\right)\right],
\end{equation}
%%%%%%%%%%%%%%%%%%%%%%%%%%%%%%%%%%%%%%%%%%%%%%%
where $\dot{M}_0$ is the mass accretion rate at $t=t_{\rm halt}$ and is about 10\% of the value expected from the usual power-law behavior at $t=t_{\rm halt}$, and $\beta_0\sim0.43$, $f_1/f_0\sim 0.5$, $f_{\rm PL}\sim 0.24$ are the parameters of the semi-analytical model of \citet{Ishizaki21a} calibrated using numerical simulations for a GW170817-like event.
According to this formula, at $t\sim3$--$4t_{\rm halt}$, the mass accretion rate decreases rapidly to about 1\% for $p=1.1$--$1.5$.
Therefore we predict that the X-ray light curve fades very steeply at least in a few decades, regardless of the duration of the super-Eddington phase.

Figure \ref{fig:halting} shows the heating rate and the conditions above which the accretion is halted.
The red line shows the simple power-law heating $\dot{q}\sim 10^{10}~{\rm erg}~{\rm g}^{-1}~{\rm s}^{-1}(t/1~{\rm day})^{-1.3}$ and the red shaded area shows the uncertainty of the heating rate of the upper and lower $0.5$ digits \citep{2020arXiv201011182B}.
Note that the uncertainty in the heating rate is shown for the $\mathcal{O}(10)$ day time scale in \citep{2020arXiv201011182B}, and that the range of uncertainties would be much broader for later times ($t\gtrsim 10~{\rm day}$) (e.g., large uncertainties in the contribution of fission, which is the predominant process in the late stage, and compositional uncertainties, see also \citet{2020arXiv201011182B}).
Furthermore, the effective heating rate applied to ejecta after $\mathcal{O}(10)$ days is several times smaller due to the decreasing thermalization efficiency of gamma rays \citep{2016MNRAS.459...35H,2016ApJ...829..110B,2018MNRAS.481.3423W,2019ApJ...876..128K,2020ApJ...891..152H} (in contrast to the fact that charged particles such as beta decay still contribute as a heating source) (see discussion in \citet{Ishizaki21a}).
The black dot with an error bar is the heating rate at $t=43$ day obtained by \citet{2019MNRAS.tmpL..14K} using the mid-infrared observations by Spitzer, which are consistent with the heating rates shown by the red lines.
Note that this data point is the intrinsic radioactive heating rate, which is corrected for the effects of thermalization efficiency.
This heating rate is estimated from data obtained in a relatively narrow band, and does not include the spectral information of the emission lines expected from kilonova/macronova emission during the nebular phase, so that there is likely to be a large systematic error.
Since the current X-ray light curve suggests the super-Eddington phase, the halting time has not likely been reached yet.
This implies a heating rate profile where the heating rate at $\mathcal{O}(1)$ day is small in the range of the shaded region, and the heating of the ejecta by gamma rays becomes inefficient around $\mathcal{O}(10)$ days.
Furthermore, it is suggested that the heating rate at a few years is $\dot q \lesssim 1.25\times 10^5 ~{\rm erg}~{\rm g}^{-1}~{\rm s}^{-1}$, which is about $10$ times below the power-law extrapolation of the heating rate necessary for the kilonova/macronova emission.
Note that this suppression factor $\sim 10$ is explained by the uncertainties of the heating rate and the gamma-ray escape as mentioned above.
According to \citet{2020arXiv201011182B}, when the fission contributes effectively, the heating rate becomes relatively large after $\mathcal{O}(10)$ days, so that our result supports a model in which the fission is not significant.

%% OLD VERSION
% \widel{
% Figure \ref{fig:halting} shows the heating rate and the conditions above which the accretion is halted.
% The red line shows the simple power-law heating $\dot{q}\sim 10^{10}~{\rm erg}~{\rm g}^{-1}~{\rm s}^{-1}(t/1~{\rm day})^{-1.3}$ .
% The black dot with an error bar is the heating rate at t=43 day obtained by \citet{2019MNRAS.tmpL..14K} using the mid-infrared observations by Spitzer, which are consistent with the heating rates shown by the red lines.
% Since the current X-ray light curve suggests the super-Eddington phase, the halting time has not likely been reached yet.
% This suggests that the heating rate at a few years is $\dot q \lesssim 5\times 10^5 ~{\rm erg}~{\rm g}^{-1}~{\rm s}^{-1}$, which is about two times below the power-law extrapolation of the heating rate necessary for the kilonova/macronova emission.
% Note that this factor two is explained by the gamma-ray escape as mentioned above.
% According to \citet{2020arXiv201011182B}, when the fission contributes effectively, the heating rate becomes relatively large after $\mathcal{O}(10)$ days, so that our result supports a model in which the fission is not significant.
% }

\section{Discussion and Conclusion} \label{sec:summary}

We have found that the year-scale X-ray excess in GW170817 can be explained by the X-ray from the fallback accretion of the post-merger ejecta (most likely disk wind).
The observed X-ray flux is consistent with that in the super-Eddington regime.
In this model, the X-ray flux will remain constant or diminish, but never increase, in contrast to the kilonova/macronova afterglow.
In general, although it depends on the temperature distribution of the accretion disk, the emission spectrum from the accretion flow is much harder than the broad non-thermal spectrum of the kilonova/macronova afterglow 
(see \citet{2017MNRAS.469.2997N,2017PASJ...69...92K,2020arXiv201205386K} for accretion flow, and see \citet{2019ApJ...886L..17H,2021arXiv210304821B} for the kilonova/macronova afterglow).
This is consistent with the fact that the radio emission of GRB 170817A does not show any re-brightening \citep{2021arXiv210304821B}.
In addition, by assuming that the accretion is still in the super-Eddington phase, we have obtained a constraint on the current radioactive heating rate.
If the heating continues as suggested by the modeling of kilonova/macronova emission, the mass accretion rate will completely fade due to the halting process shown in \citet{Ishizaki21a}, so that the super-Eddington regime can not maintain.
This suggests that the thermalization of the gamma-ray due to the radioactive decay of $r$-process elements becomes insufficient after $\mathcal{O}(10)$ day.
Furthermore, our model favors models in which the heating rate is relatively small after $\mathcal{O}(10)$ days.
According to \citet{2020arXiv201011182B}, in such models, the fission is not significant after $\mathcal{O}(10)$ days.
Future X-ray observations will further limit the uncertainty of nuclear physics.

As seen in Equation (\ref{eq:tedd_pm}) and Figure \ref{fig:lx}, if we adopt the observation- and simulation-based value of parameters (i.e, $M_{\rm ej}=0.08 \Msun$, $v = 0.04c$), the time $t_0$ when the fallback accretion starts is about $5$ sec or more.
According to \citet{2020ApJ...901..122F}, for the BH torus system with the viscosity parameter $\alpha=0.05$, the duration of the viscously driven wind is about $1$ sec.
This is characterized by the viscous timescale, which varies inversely with the viscosity parameter $\alpha$.
If $t_0=5$ sec corresponds to this viscous timescale, 
our result implies $\alpha \sim 0.01$.
Furthermore, \citet{2020PhRvD.101h3029F} showed that for an accretion disk with $\alpha=0.04$, the disk material with a mass of about $0.1 \Msun$ at the inner edge of the disk accretes to the central object in about $1$ sec after coalescence.
Since this also corresponds to the viscous timescale, a similar value $\alpha \sim 0.01$ makes the accretion last for $4$--$5$ sec.
This is consistent with the activity time of the jet driving the sGRB \citep[e.g.,][]{2020MNRAS.491.3192H}.

Assuming that the super-Eddington phase is still ongoing, we can estimate the initial fallback timescale $t_0$ by using Equation (\ref{eq:tedd_pm}), namely
%%%%%%%%%%%%%%%%%%%%%%%%%%%%%%%%%%%%%%%%%%%%%%%
\begin{equation}\label{eq:t0_obs}
t_0\sim 3.9~\sec
~\eta_{0.1}^{-2.14}
~f_{{\rm edd},2}^{2.14}
~\left(\frac{t_{\rm edd}}{1300~{\rm day}}\right)^{3.57}
~\left(\frac{M}{2.7~M_{\odot}}\right)^{1.71}
~\left(\frac{M_{\rm ej}}{0.08~M_\odot}\right)^{-2.14}
~\left(\frac{v_{\rm ej}}{0.04c}\right)^{1.29}.
\end{equation}
%%%%%%%%%%%%%%%%%%%%%%%%%%%%%%%%%%%%%%%%%%%%%%%
If we adopt phenomenological ejecta parameters $M_{\rm ej} \sim 0.08 \Msun$ and $v_{\rm ej} \sim 0.1$--$0.2c$ which are directly derived from the kilonova/macronova light curve \citep[e.g.,][]{2017ApJ...848L..19C,2017ApJ...848L..17C,2017Sci...358.1559K,2017ApJ...851L..21V} rather than simulations that take ejecta morphology into account, such as in \citet{2020arXiv201214711K}, the initial fallback time $t_0$ is required to be greater than about $30$ sec.
Furthermore, as can be seen from Equation (\ref{eq:t0_obs}), future X-ray observations will give us the opportunity to explore the properties of the accretion system on longer time scales (e.g., $t_0 \sim 100$--$1000$ sec in $10$ years of observation).
It is interesting to note that the initial fallback timescale of $t_0 \sim 30$--$10^4$ sec approximately coincides to the typical timescale of extended emission seen in sGRB \citep{2013MNRAS.430.1061R,2017ApJ...846..142K,2019ApJ...877..147K}.
More work is needed to find the connection of the post-merger ejecta in the BNS merger and sGRB \citep[e.g.,][]{2007MNRAS.376L..48R,2007NJPh....9...17L,2009MNRAS.392.1451R,2010MNRAS.406.2650M,2015ApJ...804L..16K,2017ApJ...846..142K,2019MNRAS.485.4404D}.

{\it Note added}: As this Letter was being completed, we learned of an independent study by \citet{2021arXiv210402070H}, which comes to a similar theoretical conclusion.
The excess also becomes $\sim 3.5$--$4.3 \sigma$.

\acknowledgments
% We thank for xxxxxxx valuable comments.
We thank Kyohei Kawaguchi and Sho Fujibayashi for providing us the data of the profile of the post-merger ejecta.
We also thank Kazuya Takahashi, Hamid Hamidani, Tomoki Wada, Koutarou Kyutoku, Masaru Shibata, and Takashi Nakamura for valuable comments.
We thank the participants and the organizers of the workshops with the identification number YITP-T-19-04, YKIS2019 and YITP-T-20-19 for their generous support and helpful comments.
This work is supported by Grants-in-Aid for Scientific
Research No. 18H01213 (KK), 20H01901, 20H01904, 20H00158, 18H01213, 18H01215, 17H06357, 17H06362, 17H06131 (KI) from the Ministry
of Education, Culture, Sports, Science and Technology (MEXT) of Japan.

\bibliography{sample63}{}
\bibliographystyle{aasjournal}



\end{document}